# Synthesis and characterization of super-nonazethrene

Elia Turco,[⊥,‡] Shantanu Mishra,[⊥,‡,†,*] Jason Melidonie,[∥,‡] Kristjan Eimre,[⊥] Sebastian Obermann,[∥] Carlo A. Pignedoli,[⊥] Roman Fasel,[⊥,§] Xinliang Feng,[∥,○,*] and Pascal Ruffieux[⊥,*]

[⊥]nanotech@surfaces laboratory, Empa − Swiss Federal Laboratories for Materials Science and Technology, 8600 Dübendorf, Switzerland

[∥]Faculty of Chemistry and Food Chemistry, and Center for Advancing Electronics Dresden, Technical University of Dresden, 01069 Dresden, Germany

[○]Department of Synthetic Materials and Functional Devices, Max Planck Institute of Microstructure Physics, 06120 Halle, Germany

[§]Department of Chemistry, Biochemistry and Pharmaceutical Sciences, University of Bern, 3012 Bern, Switzerland

**ABSTRACT:** Beginning with the early work of Clar et al. in 1955, zethrenes and their laterally-extended homologues, super-zethrenes, have been intensively studied in the solution phase, and are widely investigated as optical and charge transport materials. Super-zethrenes are also considered to exhibit an open-shell ground state. Zethrenes may thus serve as model compounds to investigate nanoscale π-magnetism. However, their synthesis is extremely challenging due to their high reactivity. We report here a combined in-solution and on-surface synthesis of the hitherto largest zethrene homologue – super-nonazethrene – on Au(111). Using single-molecule scanning tunneling microscopy and spectroscopy, we show that super-nonazethrene exhibits an open-shell singlet ground state featuring a large spin polarization-driven electronic gap of 1 eV. We obtain real-space maps of the frontier molecular orbitals, and find that they correspond to singly occupied molecular orbitals. In consistence with the emergence of an open-shell ground state, high-resolution tunneling spectroscopy reveals inelastic singlet-triplet spin excitations in super-nonazethrene, characterized by a strong intramolecular magnetic exchange coupling of 51 meV. Further insights are gained by mean-field and many-body perturbation theory calculations. Given the paucity of zethrene chemistry on surfaces, our results therefore provide unprecedented access to large open-shell zethrene compounds amenable to scanning probe measurements, with potential application in molecular spintronics.

Polycyclic aromatic hydrocarbons (PAHs) exhibit a range of physico-chemical properties of both fundamental interest and technological relevance, which is highly tunable through variations in molecular size, shape and edge structure. For example, while PAHs with armchair edges are generally stable with large HOMO-LUMO gaps (where HOMO and LUMO correspond to the highest occupied and lowest unoccupied molecular orbitals, respectively), PAHs with zigzag edges are more reactive and feature relatively small HOMO-LUMO gaps. PAHs with zigzag edges may also exhibit an open-shell ground state, which opens extensive opportunities in molecular electronics, non-linear optics and quantum computation.[1,2] Synthesis of open-shell PAHs in solution is challenging due to their high intrinsic reactivity, and requires kinetic and thermodynamic stabilization to confer stability for requisite characterization. Alternatively, open-shell PAHs can also be synthesized and stabilized on solid surfaces via on-surface synthesis[3] under ultra-high vacuum. This approach additionally offers the possibility to study the chemical, electronic and magnetic structures of single molecules using scanning tunneling microscopy (STM) and spectroscopy (STS).

In recent years, on-surface synthesis has been utilized to synthesize a number of PAHs with zigzag edges, such as acenes,[4–7] periacenes,[8–10] anthenes,[11] periacenoacenes,[12] triangulenes[13–19] and Clar's goblet,[20] with many of them featuring an open-shell ground state. Another prominent class of open-shell PAHs are zethrenes,[21] which are Z-shaped PAHs with mixed zigzag and armchair edges (Scheme 1). While the parent zethrene (first reported by Clar et al.[22]) is a closed-shell compound, longitudinal extension of zethrenes generates proaromatic quinodimethane moieties which tend to promote an open-shell ground state, as has been demonstrated through the syntheses of heptazethrene[23] to nonazethrene.[24] Furthermore, lateral extension of zethrenes leads to the so-called super-zethrene compounds, which are considered to be inherently open-shell.[21] While the parent super-zethrene has not been realized, syntheses of the larger homologues super-heptazethrene[25] and super-octazethrene[26] have been recently reported. In contrast to the rich chemistry of zethrenes in the solution phase, on-surface synthesis of zethrenes has remained unexplored. Only last year, we reported the on-surface synthesis of super-heptazethrene,[27] which, however, exhibited a closed-shell ground state. Access to larger zethrenes should therefore offer the possibility of investigating unconventional π-magnetism at the single molecule scale, especially in view of the potential application of zethrenes in molecular electronics and optics.[21] Here, we report a combined in-solution and on-surface synthesis of the hitherto largest zethrene compound, super-nonazethrene (**1**). Our approach furthermore allows for in situ structural, electronic and magnetic characterization of **1** on an inert Au(111) surface using bond-resolved STM[28,29] and STS, using which we demonstrate that **1** exhibits an open-shell singlet ground state with an experimental singlet-triplet gap of 51 meV.

**Scheme 1. (a) Closed-shell Kekulè and open-shell non-Kekulè resonance structures of zethrenes and super-zethrenes, and their structural relationship to the parent zethrene. (b) Combined in-solution and on-surface synthetic route toward 1.**[a,b]

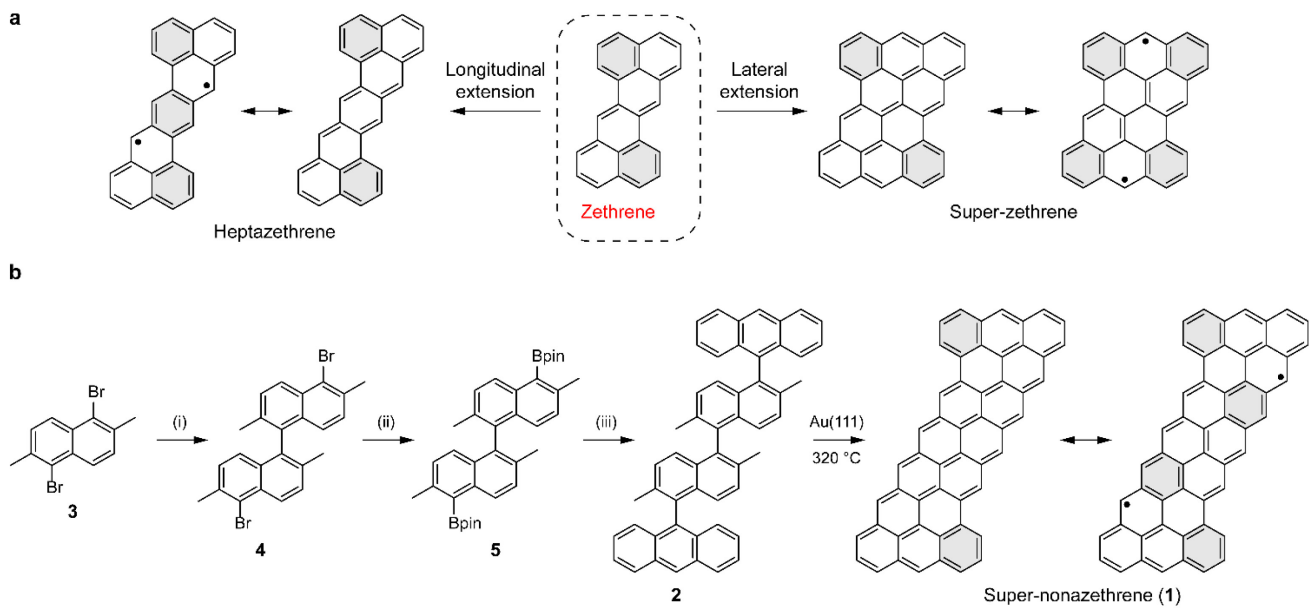

[a]Filled rings denote Clar sextets. [b]Reagents and conditions: (i) *n*-BuLi, CuCl₂, THF, −78 °C to room temperature, 18 h, 58%; (ii) KOAc, (Bpin)₂, Pd(dppf)Cl₂, DMSO, 80 °C, 20 h, 99%; and (iii) 9-bromoanthracene, K₃PO₄, Pd₂(dba)₃, DPEPhos, PhMe/EtOH/H₂O, 110 °C, 3 d, 39%.

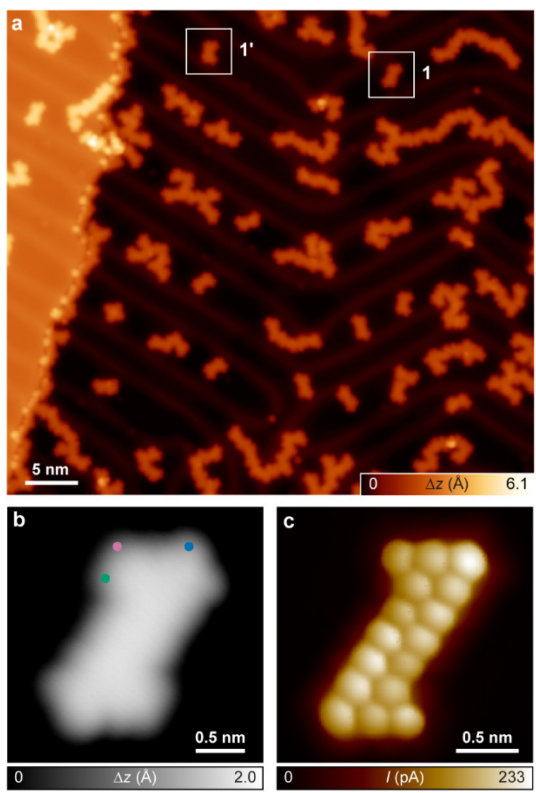

**Figure 1.** (a) Overview STM image after annealing **2** on Au(111) at 320 °C ($V$ = −0.7 V, $I$ = 70 pA). **1** and **1'** are highlighted. (b, c) High-resolution STM image (b) ($V$ = −0.1 V, $I$ = 50 pA) and corresponding bond-resolved STM image (c) ($V$ = −5 mV, $I$ = 50 pA, $\Delta h$ = −0.8 Å) of **1** on Au(111), acquired with a carbon monoxide (CO) functionalized tip.

To synthesize **1**, we designed the precursor 9,9'-(2,2',6,6'-tetramethyl-[1,1'-binaphthalene]5,5'-diyl) dianthracene (**2**), which undergoes thermally induced cyclodehydrogenation and oxidative cyclization of methyl groups on a metal surface (Scheme 1). The synthesis of **2** started with the treatment of 1,5-dibromo-2,6-dimethylnaphthalene (**3**)[30] with *n*-BuLi and CuCl₂ at −78 °C, which provided the dimer 5,5'-dibromo-2,2',6,6'-tetramethyl-1,1'-binaphthalene (**4**) in good yield.[31] In a next step, 2,2'-(2,2',6,6'-tetramethyl-[1,1'-binaphthalene]-5,5'-diyl)bis(4,4,5,5-tetramethyl-1,3,2-dioxaborolane) (**5**) was obtained under Suzuki–Miyaura conditions in quantitative yield using [1,1'-bis-(diphenylphosphino)ferrocene]dichloropalladium(II)

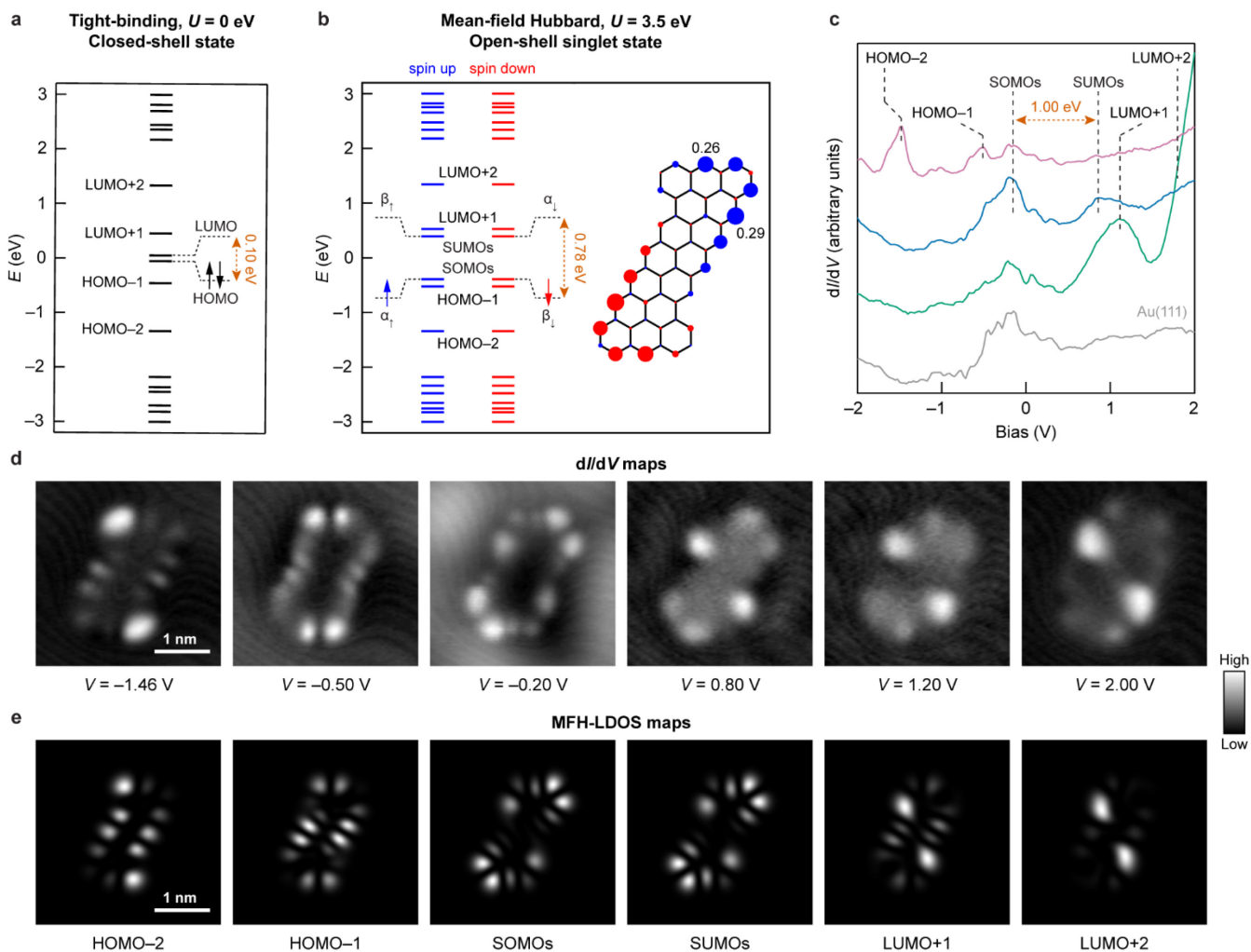

**a** Tight-binding, $U = 0$ eV
Closed-shell state

**b** Mean-field Hubbard, $U = 3.5$ eV
Open-shell singlet state

**c**

**d** d$I$/d$V$ maps

**e** MFH-LDOS maps

HOMO–2    HOMO–1    SOMOs    SUMOs    LUMO+1    LUMO+2

**Figure 2.** (a, b) TB (a) and MFH (b) energy spectrum of **1**. $U$ denotes the on-site Coulomb repulsion. Also shown is the spin polarization plot of **1**, where blue and red filled circles denote mean populations of spin up and spin down electrons, respectively, and the numbers denote the two largest mean populations of spin up electrons. (c) d$I$/d$V$ spectroscopy on **1** revealing molecular orbital resonances (open feedback parameters: $V = -2.0$ V, $I = 350$ pA; $V_{rms} = 16$ mV). Acquisition positions are indicated in Fig. 1b. (d) Constant-current d$I$/d$V$ maps at the resonances indicated in (c) ($I = 300$–320 pA; $V_{rms} = 30$ mV). (e) MFH-LDOS maps of the HOMO–2 to LUMO+2 of **1**.

dichloride as the catalyst, potassium acetate as the base and bis(pinacolato)diboron ((Bpin)$_2$) as the borylation reagent.[32] Finally, anthracene substituents were introduced to **5** through the Suzuki reaction, leading to the formation of **2** in 39% yield under the catalysis of tris(dibenzylideneacetone)dipalladium(0) (Pd$_2$(dba)$_3$) and bis[(2-diphenylphosphino)-phenyl]ether (DPEPhos).[33]

Subsequently, a submonolayer coverage of **2** was sublimed on a Au(111) surface held at room temperature, and the surface was annealed to 320 °C to promote the on-surface reactions. STM imaging of the surface after the annealing process revealed the predominance of covalently coupled molecular clusters (~80%) and a minority of individual molecules (~20%) (Fig. 1a). High-resolution STM imaging showed that ~22% of the individual molecules present a uniform Z-shaped topography consistent with **1** (Fig. 1b), which is confirmed via bond-resolved STM imaging (Fig. 1c), thereby demon-strating the success of our synthetic strategy. We note that the remaining 78% of individual molecules consist of partially reacted precursor molecules, and an isomer of **1**, referred to as **1'**, which is formed by a 180° rotation of one of the 9-(2,6-dimethylnaphthalen-1-yl)anthracene moieties of **2** relative to the other, and its subsequent cyclodehydrogenation and oxidative cyclization (Fig. S1). The low yield of **1** on Au(111) is in marked contrast with our previously reported synthesis of super-heptazethrene on Au(111)[27] with a yield of 82% under similar reaction conditions, which likely arises from the open-shell ground state of **1**, as we demonstrate below. With the aim of increasing the yield of **1**, we also explored its synthesis on the more active Ag(111) and Cu(111) surfaces (Fig. S2), where however the absolute yields of **1** remained below 10%.

To study the electronic structure of **1**, we start by performing nearest-neighbor tight-binding (TB) calculations, which do not involve electron-electron interac tions necessary to describe magnetism, and therefore lead to a closed-shell state of **1**. Figure 2a shows the TB energy spectrum of **1**, where the principal features are a pair of frontier states located close to zero energy that correspond to the HOMO and LUMO, with the corresponding wave functions localized at the zigzag termini of **1** (Fig. S3 and S4). The calculated HOMO-LUMO gap of **1** is 0.10 eV. Next, we perform mean-field Hubbard (MFH) calculations accounting for electron-electron interactions, which leads to spin polarization of **1**. The frontier electronic structure is now characterized by a pair of singly occupied molecular orbitals (SOMOs) α and β (Fig. 2b), with an antiferromagnetic correlation of the populating spins (that is, an open-shell singlet ground state, see Fig. S5), in accordance with Ovchinnikov's rule.[34] The calculated wave functions of SOMOs and the associated unoccupied orbitals (SUMOs) exhibit a largely disjoint character, as is expected for an open-shell singlet diradicaloid (Fig. S6 and S7). Spin polarization of **1** results in a significant opening of the frontier gap with the calculated SOMO-SUMO gap being 0.78 eV.

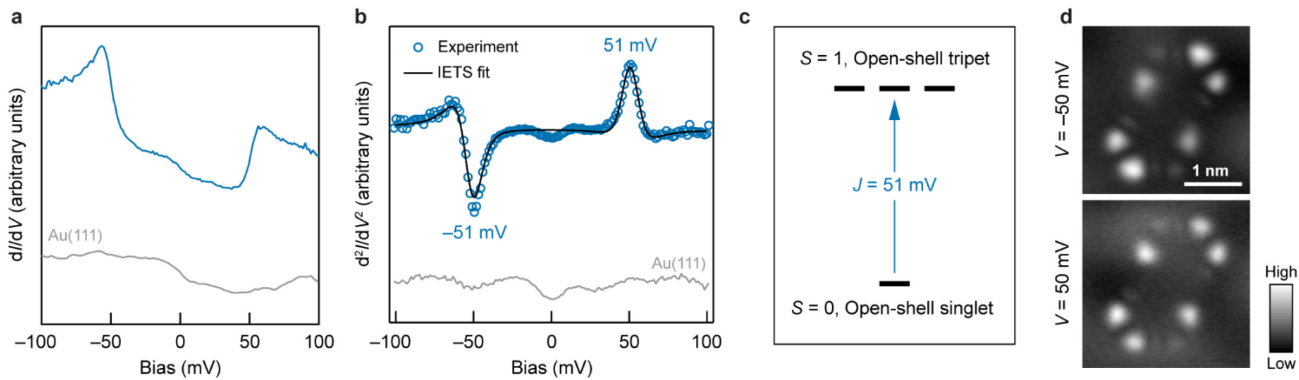

**Figure 3.** (a, b) d$I$/d$V$ (a) and d$^2I$/d$V^2$ (b) spectrum acquired on **1**, revealing singlet-triplet spin excitations (open feedback parameters: $V = -100$ mV, $I = 1.2$ nA (a) and 2.0 nA (b); $V_{rms} = 400$ µV (a) and 4 mV (b)). Acquisition position is indicated in Fig. 1b. Also shown is an inelastic electron tunneling spectroscopy (IETS) fit to the d$^2I$/d$V^2$ spectrum using a spin-½ Heisenberg dimer model, which yields the spin excitation threshold as 51 mV. (c) Schematic representation of the spin excitation. (d) Constant-current d$I$/d$V$ maps near the spin excitation threshold ($I = 280$ pA; $V_{rms} = 4$ mV).

We employed STS to probe the experimental electronic structure of **1**. d$I$/d$V$ spectroscopy on **1** (where $I$ and $V$ denote the tunneling current and bias voltage, respectively) reveals six distinguishable features in the local density of states (LDOS) at −1.46 V, −0.50 V, −0.20 V, 0.80 V, 1.20 V and around 1.80 V (Fig. 2c). Spatially resolved d$I$/d$V$ maps at −1.46 V, −0.50 V, 1.20 V and 2.00 V (Fig. 2d) exhibit an excellent match with the calculated MFH-LDOS maps of the HOMO−2, HOMO−1, LUMO+1 and LUMO+2 of **1**, respectively (Fig. 2e). Furthermore, the frontier orbital d$I$/d$V$ maps at −0.20 V and 0.80 V show similar shapes and LDOS symmetries, which suggests that at −0.20 V/0.80 V, electron tunneling from/to singly occupied orbitals are involved. The d$I$/d$V$ maps at −0.20 V and 0.80 V agree well with the MFH-LDOS maps of the SOMOs and SUMOs of **1**, which supports an open-shell ground state of **1**. In addition, we note that the frontier electronic gap of **1** on

Au(111) is 1.00 eV, which is much larger than the HOMO-LUMO gap of 0.23 eV of super-heptazethrene on Au(111).[27] Given that the HOMO-LUMO gap of a homologous series of molecules should decrease with increasing molecular size, the counterintuitive increase of the frontier electronic gap of **1** compared to its smaller homologue super-heptazethrene therefore cannot be explained by a closed-shell ground state of **1**. This strongly suggests that the electronic gap of **1** corresponds to a spin polarization-driven Coulomb gap and the frontier states correspond to SOMOs. Our interpretation is supported by spin-polarized density functional theory calculations on **1**, which yield the open-shell singlet as the ground state of **1**, with the open-shell triplet and the closed-shell states 53 meV and 0.12 eV higher in energy, respectively (Fig. S8). Furthermore, we performed eigenvalue self-consistent GW calculations including screening effects from the underlying surface (that is, GW+IC, where IC denotes image charge), for the open-shell singlet and closed-shell states of **1** (Fig. S8). From our calculations, the experimental frontier gap of 1.00 eV is consistent with the GW+IC Coulomb gap of 0.70 eV for the open-shell singlet state of **1**, while clearly disagreeing with the GW+IC HOMO-LUMO gap of 90 meV for the closed-shell state of **1**, which thus confirms the open-shell ground state of **1** on Au(111).

A final and direct experimental proof of the open-shell ground state of **1** comes from low-bias d$I$/d$V$ spectroscopy on **1** that reveals conductance steps symmetric about the Fermi energy, which correspond to inelastic excitations[35] (Fig. 3a,b). Given that their occurrence is concurrent with the observation of SOMOs, we ascribe these excitations to singlet-triplet spin excitations (Fig. 3c).[20,36–38] In this picture, tunneling electrons may cause spin excitation of a coupled spin system whenever their energy e$V \geq J$, where $J$ denotes the magnetic exchange coupling of the radicals in **1**, or equivalently, the singlet-triplet gap of **1**. The excitation threshold, extracted from a fit to the experimental d$^2I$/d$V^2$ spectrum with a spin-½ Heisenberg dimer model,[38] amounts to 51 meV (Fig.

3b). The spatial distribution of the spin excitation signal is visualized through acquisition of d$I$/d$V$ maps around the spin excitation threshold (Fig. 3d), which resemble the SOMOs/SUMOs LDOS.

In summary, we have successfully synthesized unsubstituted and stable super-nonazethrene on Au(111). We employ STM and STS measurements at submolecular resolution to probe the chemical, electronic and magnetic structure of **1**. Comparison of our experimental STS data with theoretical calculations provides conclusive evidence of the open-shell singlet ground state of **1**. Low-bias STS measurements reveal singlet-triplet spin excitations in **1**, with the corresponding singlet-triplet gap being 51 meV, exceeding the room temperature thermal energy (25.7 meV at 298 K) by almost a factor of two. Our results provide crucial insights into the emergence of robust π-magnetism in PAHs, with implications in low-dimensional magnetism and organic spintronics.

## ASSOCIATED CONTENT

**Supporting Information**. Detailed synthetic description of chemical compounds reported in this study and associated solution characterization data, additional STM and STS data, additional calculations, and experimental and calculation methods. This material is available free of charge via the Internet at http://pubs.acs.org.

## AUTHOR INFORMATION

**Corresponding Author**

* shantanu.mishra@empa.ch
* xinliang.feng@tu-dresden.de
* pascal.ruffieux@empa.ch

**Present Addresses**

† IBM Research – Zurich, 8803 Rüschlikon, Switzerland

**Author Contributions**

‡ These authors contributed equally

**Notes**

The authors declare no competing financial interests. The source data for the results presented in this work is publicly accessible from the Materials Cloud platform (DOI: 10.24435/materialscloud:j7-51).

## ACKNOWLEDGMENT

This work was supported by the Swiss National Science Foundation (grant no. 200020-182015 and IZLCZ2-170184), the NCCR MARVEL funded by the Swiss National Science Foundation (grant no. 51NF40-182892), the EU Horizon 2020 research and innovation program – Marie Skłodowska-Curie grant no. 813036 and Graphene Flagship Core 3 (grant no. 881603), the Office of Naval Research (grant no. N00014-18-1-2708), ERC Consolidator grant (T2DCP, no. 819698), and the German Research Foundation within the Cluster of Excellence – Center for Advancing Electronics Dresden and EnhanceNano (grant no. 391979941). Computational support from the Swiss National Supercomputing Centre under project ID s904 is acknowledged.

Table of Contents artwork

Super-nonazethrene

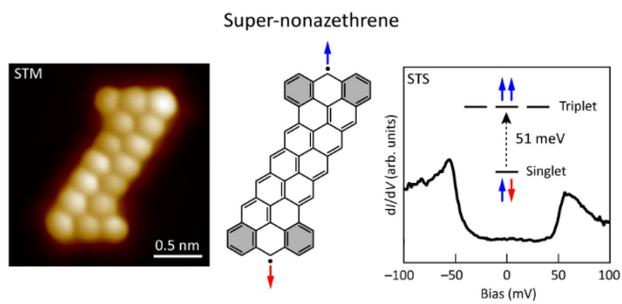

## Supporting Information

# Synthesis and characterization of super-nonazethrene

Elia Turco, Shantanu Mishra, Jason Melidonie, Kristjan Eimre, Sebastian Obermann, Carlo A. Pignedoli, Roman Fasel, Xinliang Feng, and Pascal Ruffieux

Contents

# 1. STM/STS data and calculations

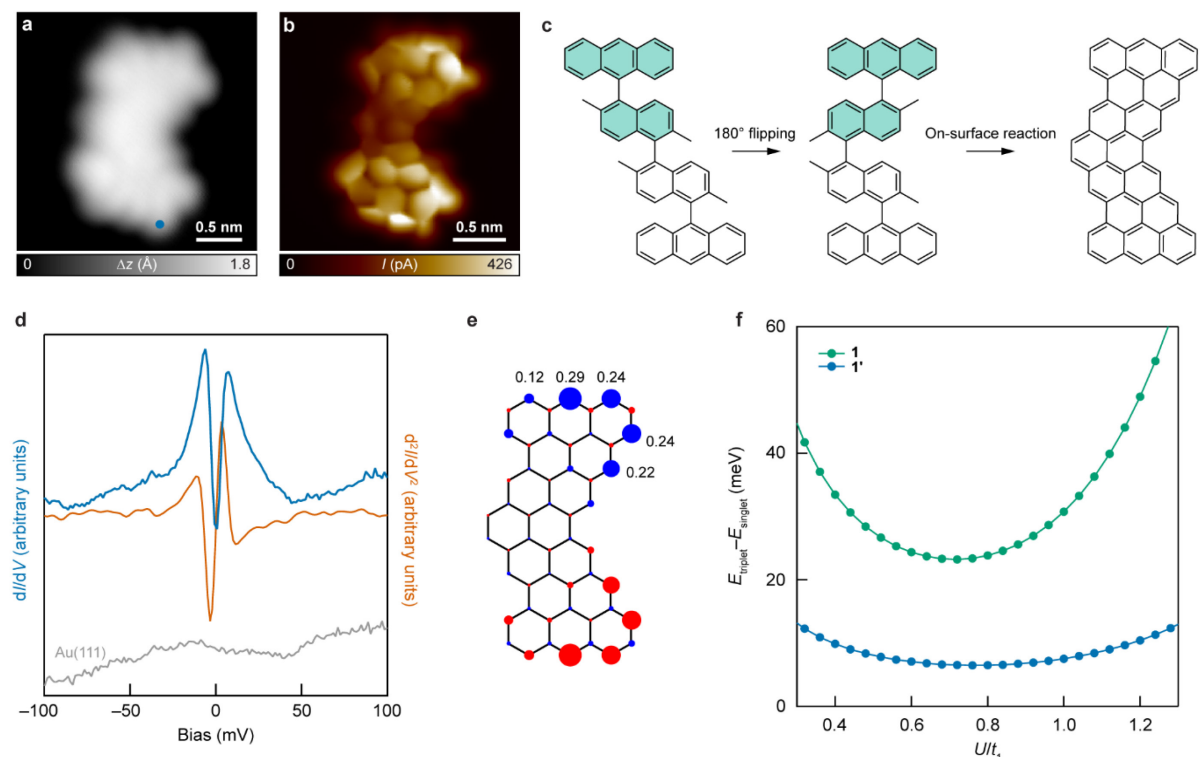

**Figure S1.** (a, b) High-resolution STM image (a) ($V = -0.1$ V, $I = 50$ pA) and corresponding bond-resolved STM image (b) ($V = -5$ mV, $I = 50$ pA, $\Delta h = -0.8$ Å) of the isomer **1'** on Au(111), acquired with a CO functionalized tip. (c) Schematic illustration of the formation of **1'** from **2**, involving a 180° flipping of one of the 9-(2,6-dimethylnaphthalen-1-yl)anthracene moieties (filled rings) with respect to the other, and subsequent cyclodehydrogenation and oxidative cyclization. (d) d$I$/d$V$ spectrum acquired on **1'**, revealing singlet-triplet spin excitations with an exchange coupling of 3 meV (open feedback parameters: $V = -100$ mV, $I = 500$ pA; $V_{rms} = 400$ µV). Also shown is the corresponding d²$I$/d$V$² spectrum obtained from numerical differentiation (note: a binomial smoothing with 5 iterations was applied to the d$I$/d$V$ spectrum before differentiation). Acquisition position is indicated in (a). (e) MFH spin polarization plot of **1'**. Blue and red filled circles denote mean populations of spin up and spin down electrons, respectively. The numbers denote the four largest mean populations of spin up electrons at the upper terminus of **1'**. (f) MFH singlet-triplet gaps of **1** and **1'** as a function of $U$ (scaled here with respect to $t_1$), for nearest-neighbor hopping. The singlet-triplet gap is calculated as the mean-field energy difference between the converged electronic structures of the open-shell triplet and the open-shell singlet configurations. Over the entire range of $U$ considered here, the open-shell singlet state is lower in energy than the open-shell triplet state for both **1** and **1'**. Furthermore, the singlet-triplet gap of **1'** is much smaller than that of **1**, which is consistent with experimental observations.

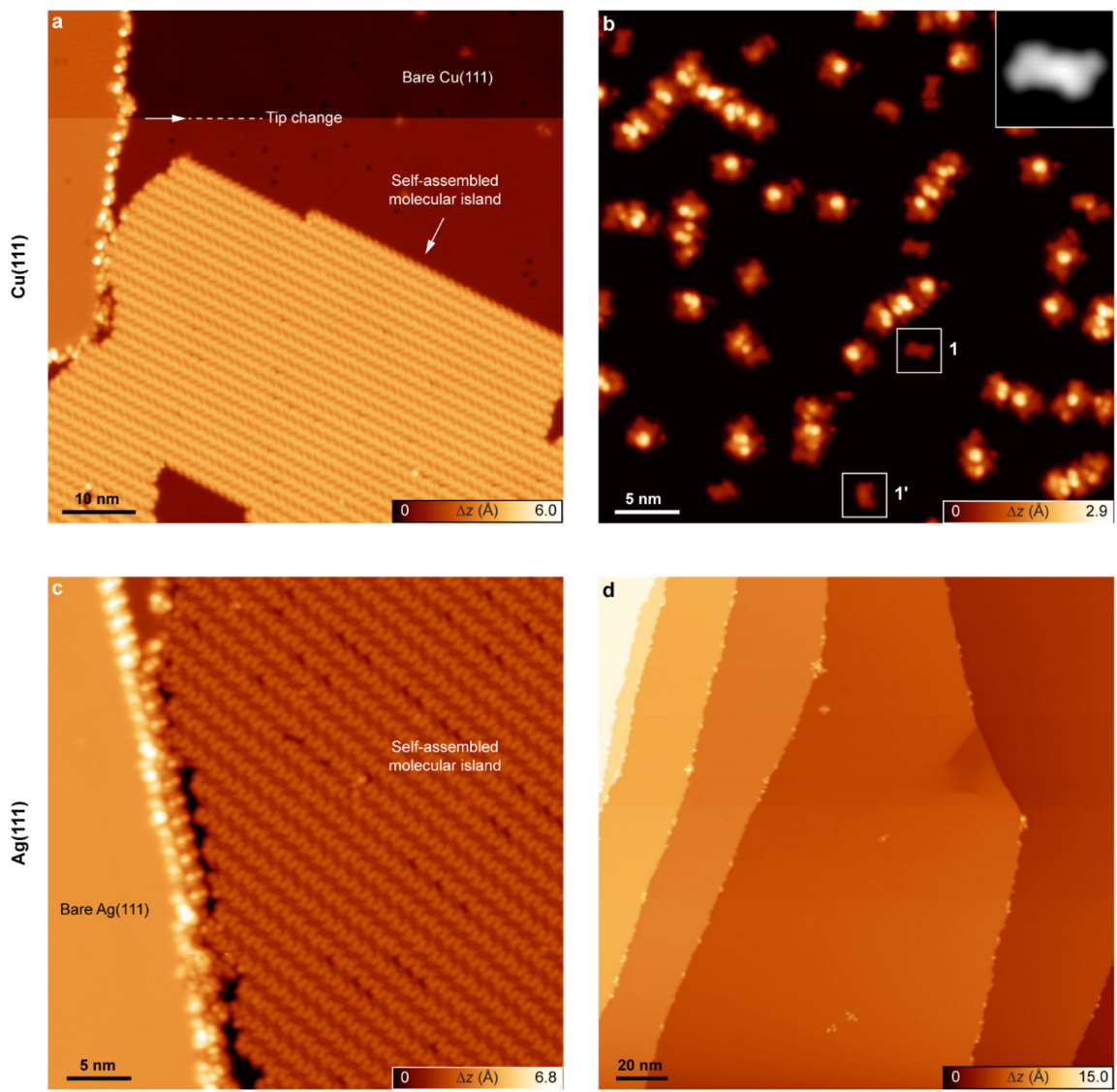

**Figure S2.** (a) Overview STM image after deposition of **2** on Cu(111) held at room temperature ($V = -1.2$ V, $I = 40$ pA). (b) Overview STM image after annealing the sample shown in (a) to 280 °C ($V = -0.7$ V, $I = 50$ pA). The surface is dominated by covalently coupled molecular clusters, with a minority of **1** and **1'** species (highlighted). High-resolution STM image of **1** on Cu(111) is shown in the inset ($V = -1.0$ V, $I = 300$ pA). (c) Overview STM image after deposition of **2** on Ag(111) held at room temperature ($V = -1.0$ V, $I = 30$ pA). (d) Overview STM image after annealing the sample shown in (a) to 300 °C ($V = -1.5$ V, $I = 30$ pA). The surface is largely devoid of molecules, indicating the **2** desorbs from Ag(111) upon annealing.

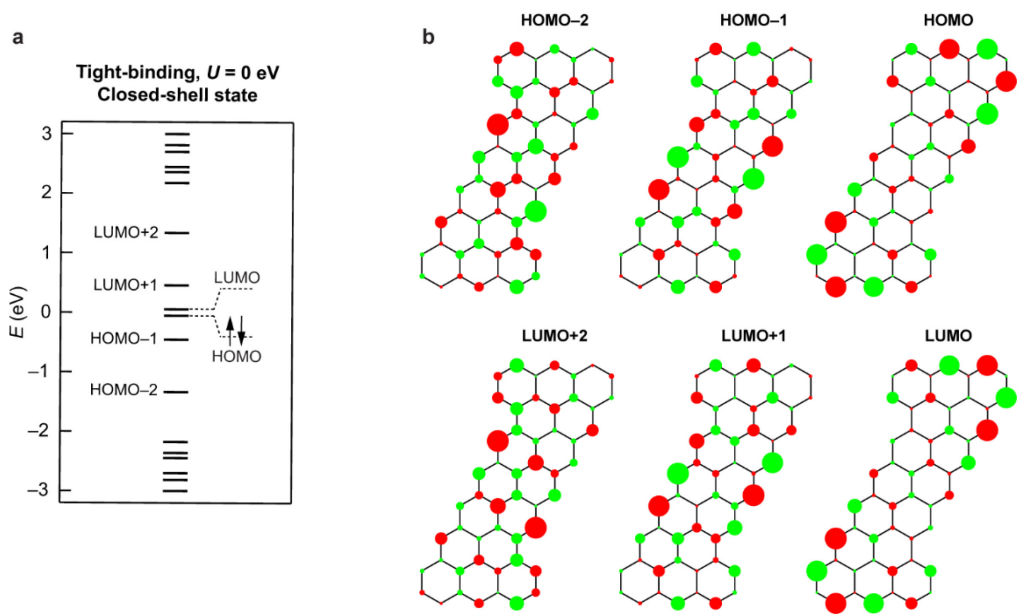

**Figure S3.** (a) Nearest-neighbor TB energy spectrum of **1**. HOMO−2 to LUMO+2 orbitals are marked. (b) TB wave functions of HOMO−2 to LUMO+2 of **1**. Red/green filled circles denote opposite phases of the wave function, while the size of the circles denotes amplitude of the wave function. The HOMO and LUMO wave functions are localized at the zigzag termini.

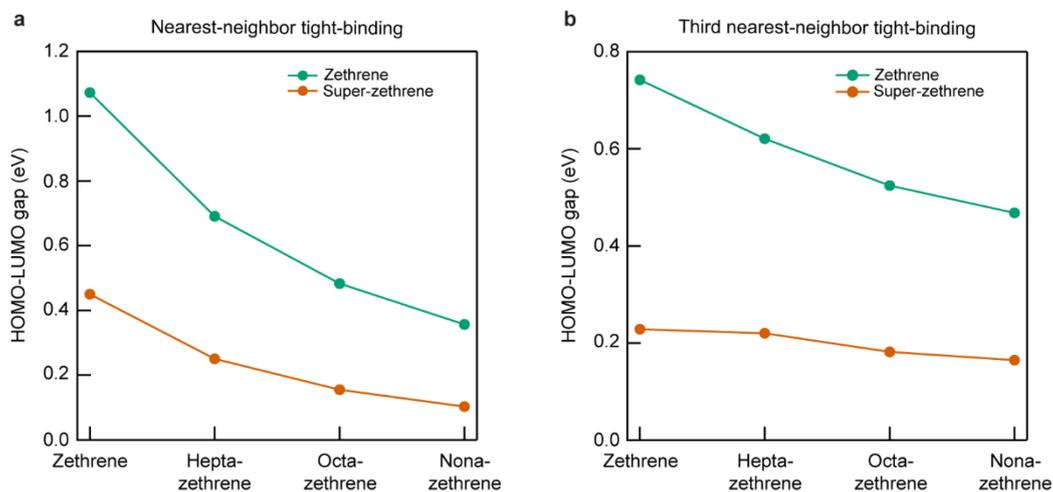

**Figure S4.** (a, b) Nearest-neighbor (a) and third-nearest-neighbor (b) TB HOMO-LUMO gaps of zethrenes and super-zethrenes. Compared to the corresponding zethrenes, super-zethrenes exhibit a much smaller HOMO-LUMO gap, which is consistent with a larger open-shell character of super-zethrenes.[1]

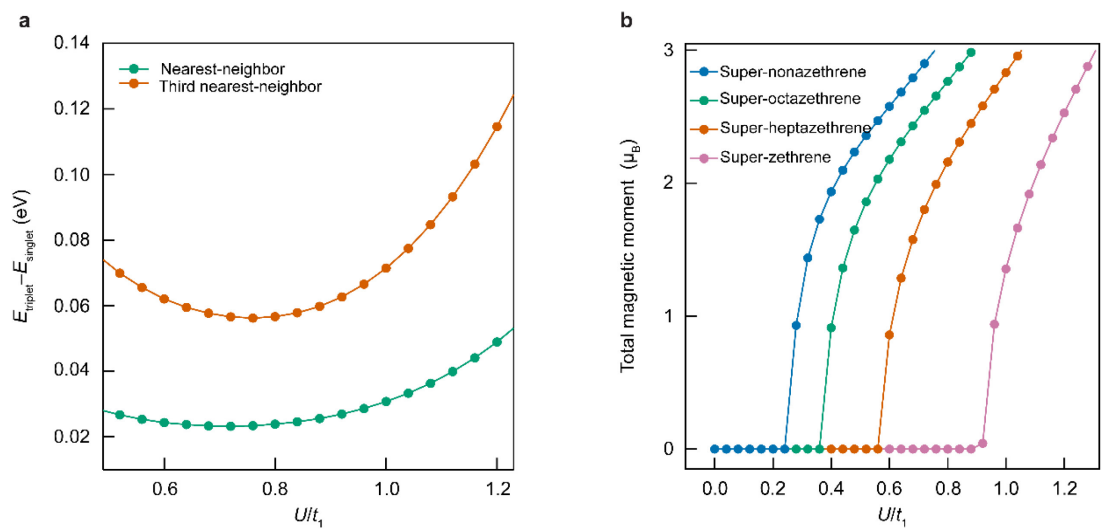

**Figure S5.** (a) MFH singlet-triplet gap of **1** as a function of $U$ (scaled here with respect to $t_1$), for both nearest-neighbor and third-nearest-neighbor hopping. Over the entire range of $U$ considered here, the open-shell singlet state is lower in energy than the open-shell triplet state. (b) Total absolute magnetic moment of the open-shell singlet states of super-zethrene, super-heptazthrene, super-octazethrene and super-nonazethrene as a function of $U$, for nearest-neighbor hopping. With increasing size, super-zethrenes are driven into an open-shell state at progressively lower $U$ values.

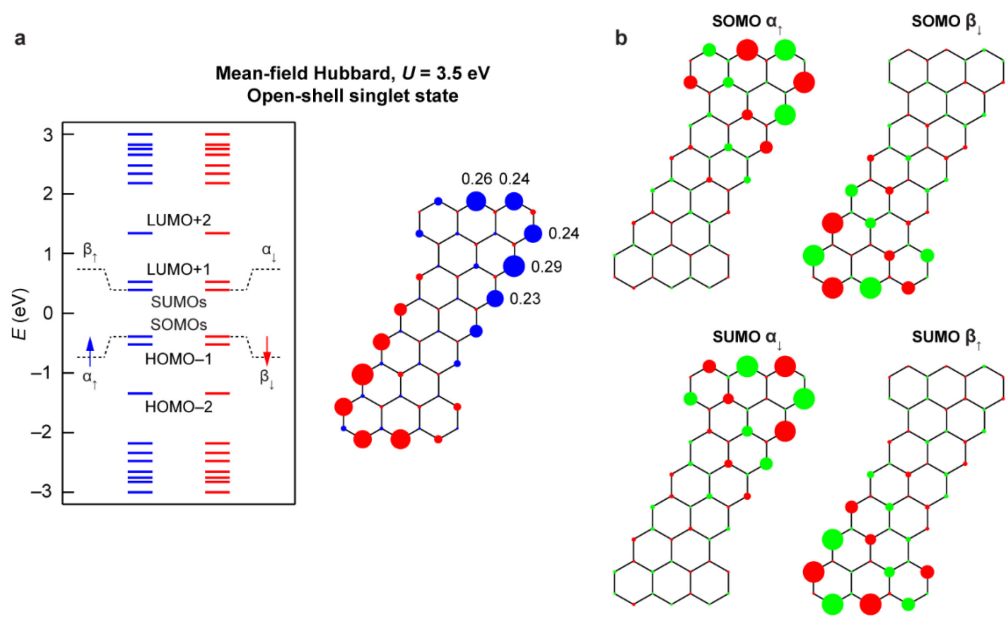

**Figure S6.** (a) Nearest-neighbor MFH energy spectrum of the open-shell singlet state of **1**. HOMO−2 to LUMO+2 orbitals are marked. $\alpha_{\uparrow(\downarrow)}$ and $\beta_{\uparrow(\downarrow)}$ denote the spin-polarized singly occupied molecular orbitals. Also shown is the spin polarization plot of **1**. (b) MFH wave functions of the SOMOs and SUMOs of **1**. The wave functions of SOMOs/SUMOs are spatially separated and localized at the opposite ends of **1**.

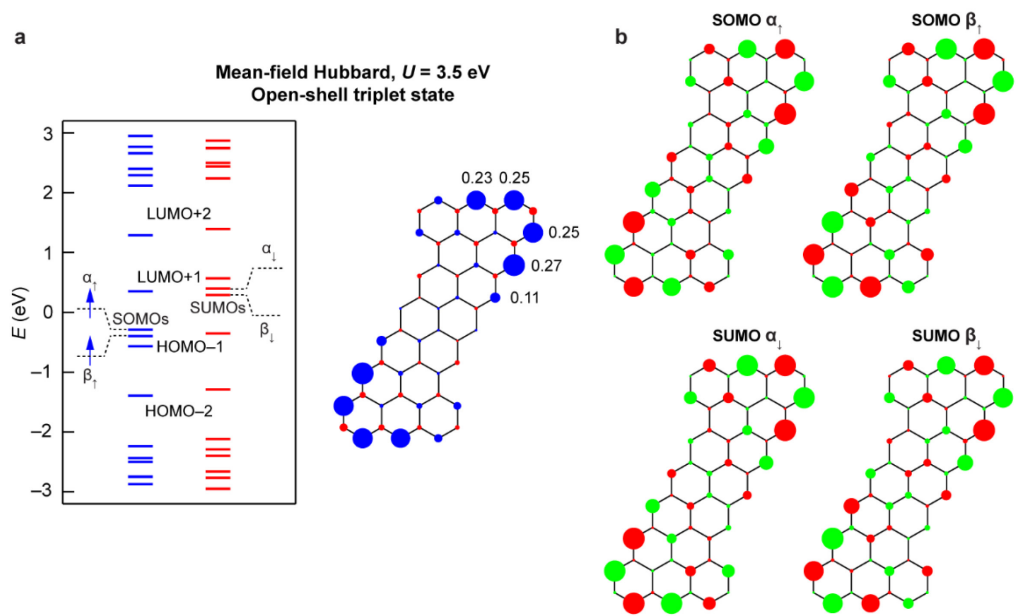

**Figure S7.** (a) Nearest-neighbor MFH energy spectrum of the open-shell triplet state of **1**. HOMO−2 to LUMO+2 orbitals are marked. Also shown is the spin polarization plot of **1**. (b) MFH wave functions of the SOMOs and SUMOs of **1**.

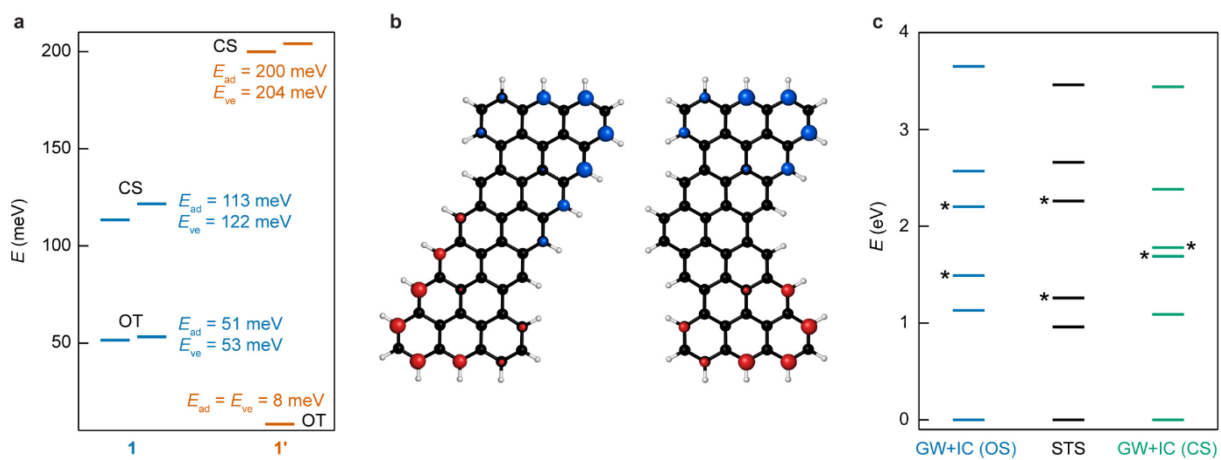

**Figure S8.** (a) Relative gas-phase energies of the open-shell triplet (OT) and closed-shell (CS) states of **1** and **1'** with respect to their open-shell singlet (OS) ground states from spin-polarized density functional theory (DFT) calculations. $E_{ad}$ and $E_{ve}$ denote the adiabatic and vertical energies, respectively. As seen experimentally and reflected in our MFH calculations (Fig. S1), the singlet-triplet gap of **1** is much larger than that of **1'**. (b) Gas-phase DFT spin density distributions of **1** and **1'** (isovalue: ±0.01 $e$/Bohr³). Blue and red isosurfaces denote spin up and spin down densities, respectively. (c) Energies of the HOMO−2 to LUMO+2 resonances of **1** on Au(111) from GW+IC calculations for the OS and CS states, and STS measurements. Corresponding frontier orbitals (HOMO/SOMOs and LUMO/SUMOs) are indicated with asterisk. Energies of the calculated and experimental levels are aligned at HOMO−2. The spin up and spin down GW+IC levels for the OS state are energetically degenerate.

## 2. Methods

### 1. Sample preparation and STM/STS measurements.

STM measurements were performed with a commercial low-temperature STM from Scienta Omicron operating at a temperature of 4.5 K, and base pressure below $5\times10^{-11}$ mbar. Au(111), Ag(111) and Cu(111) single crystal surfaces were prepared by iterative $Ar^+$ sputtering and annealing cycles. Prior to sublimation of molecules, the surface quality was checked through STM imaging. Precursor **2** was contained in a quartz crucible in powder form, and sublimed at 285 °C from a home-built evaporator onto the single crystal surfaces held at room temperature. Typical sample annealing times post deposition of precursor molecules ranged between 3 to 10 min. STM images were acquired both in constant-current (overview and high-resolution imaging) and constant-height (bond-resolved imaging) modes, $dI/dV$ spectra were acquired in constant-height mode and $dI/dV$ maps were acquired in constant-current mode. Indicated bias voltages are provided with respect to sample. Unless otherwise noted, all measurements were performed with metallic tips. $dI/dV$ spectra and maps were obtained with a lock-in amplifier (SR830, Stanford Research Systems) operating at a frequency of 860 Hz. Modulation voltages (root mean square amplitude) for each measurement is provided in the respective figure caption. Bond-resolved STM images were acquired in constant-height mode with CO functionalized tips, at biases close to the Fermi energy, and the current signal was recorded. Open feedback parameters on the molecular species and subsequent lowering of the tip height ($\Delta h$) for each image are provided in the respective figure caption. The data were processed with Wavemetrics Igor Pro software.

### 2. Tight-binding and mean-field Hubbard calculations.

TB/MFH calculations have been performed by numerically solving the mean-field Hubbard Hamiltonian with nearest-neighbor or third-nearest-neighbor hopping

$$\hat{H}_{MFH} = \sum_j \sum_{\langle \alpha,\beta\rangle_j,\sigma} -t_j c_{\alpha,\sigma}^\dagger c_{\beta,\sigma} + U \sum_{\alpha,\sigma} \langle n_{\alpha,\sigma}\rangle n_{\alpha,\bar{\sigma}} - U \sum_\alpha \langle n_{\alpha,\uparrow}\rangle \langle n_{\alpha,\downarrow}\rangle. \tag{S1}$$

Here, $c_{\alpha,\sigma}^\dagger$ and $c_{\beta,\sigma}$ denote the spin selective ($\sigma \in \{\uparrow,\downarrow\}$ with $\bar{\sigma} \in \{\downarrow,\uparrow\}$) creation and annihilation operator at sites $\alpha$ and $\beta$, $\langle \alpha,\beta\rangle_j$ ($j = \{1, 2, 3\}$) denotes the nearest-neighbor, second-nearest-neighbor and third-nearest-neighbor sites for $j = 1$, 2 and 3, respectively, $t_j$ denotes the corresponding hopping parameters (with $t_1 = 2.7$ eV, $t_2 = 0.1$ eV and $t_3 = 0.4$ eV for nearest-neighbor, second-nearest-neighbor and third-nearest-neighbor hopping[2]), $U$ denotes the on-site Coulomb repulsion, $n_{\alpha,\sigma}$ denotes the number operator, and $\langle n_{\alpha,\sigma}\rangle$ denotes the mean occupation number at site $\alpha$. Orbital electron densities, $\rho$, of the $n$th-eigenstate with energy $E_n$ have been simulated from the corresponding state vector $a_{n,i,\sigma}$ by

$$\rho_{n,\sigma}(\vec{r}) = \left| \sum_i a_{n,i,\sigma}\phi_{2p_z}(\vec{r} - \vec{r}_i) \right|^2, \tag{S2}$$

where $i$ denotes the atomic site index and $\phi_{2p_z}$ denotes the Slater $2p_z$ orbital for carbon.

### 3. Density functional theory and GW calculations.

Gas-phase DFT calculations for different spin multiplicities were performed with the Gaussian software package.[3] The PBE exchange-correlation functional[4] was used in the restricted (for the closed-shell state) and unrestricted (for the open-shell states) Kohn-Sham formulations. For geometry optimizations and single point energies, the 6-311G** and the 6-311+G** basis sets, respectively, were used.

The adsorption geometries of **1** and **1'** on Au(111) were calculated with the AiiDAlab[5] and CP2K[6] software packages. We used the PBE functional together with the DFT-D3 van der Waals corrections[7] and norm-conserving GTH pseudopotentials.[8] A TZV2P MOLOPT basis set[9] was used for C and H species, and a DZVP MOLOPT basis set for the Au species, together with a cutoff of 600 Ry for the plane wave basis set. The surface/adsorbate system was modeled within the repeated slab scheme, with a simulation cell containing 4 atomic layers of Au along the [111] direction and a layer of hydrogen atoms to suppress one of the two Au(111) surface states. 40 Å of vacuum was included in the simulation cell to decouple the system from its periodic replicas in the direction perpendicular to the surface. The Au(111) surface was modeled by a supercell of 41.27 Å × 40.85 Å corresponding to 224 surface units. Unrestricted Kohn-Sham approach was used together with an anti-ferromagnetic spin guess to model the magnetic ground state. The adsorption geometry was optimized by keeping the positions of the bottom two layers of the slab fixed to the ideal bulk coordinates, while all the other atoms were relaxed until forces were lower than 0.005 eV/Å.

The eigenvalue self-consistent GW calculations[10] were performed with the CP2K code and AiiDA[11] work-chains on the isolated molecular geometry corresponding to the adsorption conformation. The calculations were performed based on the restricted (for the closed-shell state) and the unrestricted (for the open-shell singlet state) DFT PBE wavefunctions using the GTH pseudopotentials and analytic continuation with a two-pole model. An aug-cc-pVDZ basis set[12,13] was used together with the corresponding aug-cc-pVDZ-RIFIT resolution of identity basis set.[14] To account for screening by the metal surface, we applied the image charge model by Neaton et al.,[15] and to determine the image plane position with respect to the molecular geometry, we used a distance of 1.42 Å between the image plane and the first surface layer, as reported by Kharche et al.[16]

## 4. General methods and materials in solution synthesis.

Unless otherwise stated, commercially available starting materials, catalysts, reagents, and dry solvents were used without further purification. Reactions were performed using standard vacuum-line and Schlenk techniques. All the starting materials were obtained from Sigma Aldrich, TCI, abcr, Alfa Aesar, Acros Organics, Sigma Aldrich or chemPUR. Catalysts were purchased from Strem.

Column chromatography was performed on silica ($SiO_2$, particle size 0.063–0.200 mm, purchased from VWR) or aluminum oxide ($Al_2O_3$, Alox 90 active neutral, activity stage I, purchased from Merck KGaA). Silica-coated aluminum sheets with a fluorescence indicator (TLC silica gel 60 $F_{254}$, purchased from Merck KGaA) were used for thin-layer chromatography.

NMR data were recorded on a Bruker Ascend 300 spectrometer operating at 300 MHz for $^1H$ and 76 MHz for $^{13}C$ with standard Bruker pulse programs at room temperature (296 K). Chemical shifts were referenced to $\delta_{TMS} = 0.00$ ppm ($^1H$, $^{13}C$). Chemical shifts ($\delta$) are reported in ppm. Coupling constants ($J$) are reported in Hz. Chloroform-d1 ($\delta$($^1H$) = 7.26 ppm, $\delta$($^{13}C$) = 77.16 ppm) was used as solvent. The following abbreviations are used to describe peak patterns as appropriate: s = singlet, d = doublet, and m = multiplet. Dichloromethane-d2 (99.9 atom% D) was purchased from Eurisotop.

MALDI-TOF spectra were recorded on a Bruker Autoflex Speed MALDI-TOF MS (Bruker Daltonics, Bremen, Germany). All of the samples were prepared by mixing the analyte and the matrix, 1,8-dihydroxyanthracen-9(10*H*)-one (dithranol, purchased from Fluka Analytical, purity >98%) or *trans*-2-[3-(4-*tert*-butylphenyl)-2-methyl-2-propenylidene]malononitrile (DCTB, purchased from Sigma Aldrich, purity >99%) in the solid state.

High-resolution ESI and APCI mass spectra were recorded with an Agilent 6538 UHD accurate-mass Q-TOF LC-MS system using the positive and negative mode.

Melting points were determined on a Büchi Melting Point M-560 in a range of 50–400 °C with a temperature rate of 10 °C min$^{-1}$.

FT-IR spectrometry was carried out on a Bruker Tensor II with an ATR crystal at room temperature.

# 3. Synthetic procedures

### 3.1. Synthesis of 1,5-dibromo-2,6-dimethylnaphthalene (**3**).[17]

In a 100 mL two-necked flask, 2,6-dimethylnaphthalene (10.00 g, 64.01 mmol, 1.00 eq) was dissolved in dichloromethane (30 mL) and cooled down to 0 °C. Bromine (7.21 mL, 140.82 mmol, 2.20 eq) in dichloromethane (20 mL) was added in one hour via dropping funnel. The reaction mixture was stirred at room temperature for 3 hours and quenched with a saturated aqueous $Na_2S_2O_3$ solution (30 mL). The aqueous phase was extracted three times with chloroform (100 mL) and the combined organic layers were dried over $MgSO_4$, filtered and all volatiles were evaporated under reduced pressure. The residue was recrystallized from ethyl acetate and **3** (18.60 g, 59.23 mmol) was obtained as colorless needles in a yield of 93%.

[1]H NMR (300 MHz, $CDCl_3$) $\delta$ = 8.18 (d, $J$ = 8.7 Hz, 2H), 7.40 (d, $J$ = 8.7 Hz, 2H), 2.62 (s, 6H) ppm.

[13]C NMR (76 MHz, $CDCl_3$) $\delta$ = 136.0, 132.1, 129.9, 126.5, 124.1, 24.1 ppm.

HR-MS (APCI, pos) $m/z$: [M⁺] calculated for $C_{12}H_{10}Br_2$: 311.9144, found: 311.9143, error: -0.32 ppm.

$R_f$: 0.73 (*iso*-hexane).

### 3.2. Synthesis of 5,5'-dibromo-2,2',6,6'-tetramethyl-1,1'-binaphthalene (**4**).[18]

In a 50 mL Schlenk flask, **3** (509 mg, 1.62 mmol, 1.00 eq) was dissolved in tetrahydrofuran (12 mL) and cooled down to −78 °C. *n*-Butyllithium (1.1 mL, 1.6 M in *n*-hexane, 1.10 eq) was added dropwise and stirred for 1 hour at −78 °C. $CuCl_2$ (283 mg, 2.11 mmol, 1.30 eq) was added in one portion and the reaction mixture was stirred at room temperature overnight. Water (20 mL) was added and the reaction mixture was extracted with dichloromethane (50 mL) three times. The organic phase was dried over $MgSO_4$ and filtered off. The crude product was purified by column chromatography (silica gel, *n*-hexane) and **4** (219 mg, 0.94 mmol) was obtained as white solid in 58% yield.

[1]H NMR (300 MHz, $CDCl_3$) $\delta$ = 8.35 (d, $J$ = 8.8 Hz, 2H), 7.58 (d, $J$ = 8.8 Hz, 2H), 7.07 (d, $J$ = 8.6 Hz, 2H), 6.87 (d, $J$ = 8.6 Hz, 2H), 2.57 (s, 6H), 2.01 (s, 6H) ppm.

[13]C NMR (76 MHz, $CDCl_3$) $\delta$ = 135.3, 135.2, 134.5, 132.4, 131.5, 130.3, 129.1, 126.8, 125.0, 124.5, 24.1, 19.9 ppm.

HR-MS (MALDI-TOF, dithranol) $m/z$: [M⁺] calculated for $C_{24}H_{20}Br_2$: 465.9926, found: 465.9923, error = -0.64 ppm.

$R_f$ = 0.43 (*iso*-hexane)

ATR-IR (cm⁻¹) $v$: 3056, 3004, 2952, 2918, 2852, 1592, 1487, 1463, 1435, 1376, 1347, 1316, 1209, 1191, 1167, 1141, 1029, 976, 962, 924, 890, 873, 808, 772, 735, 698, 673, 634, 610, 584, 549, 505, 471.

Mp: 183.8 °C.

### 3.3. Synthesis of 2,2'-(2,2',6,6'-tetramethyl-[1,1'-binaphthalene]-5,5'-diyl)bis(4,4,5,5-tetramethyl-1,3,2-dioxaborolane) (**5**).[19]

In a 25 mL pressure tube, **4** (152 mg, 0.32 mmol, 1.00 eq), potassium acetate (255 mg, 2.60 mmol, 8.00 eq) and 4,4,4',4',5,5,5',5'-octamethyl-2,2'-bi(1,3,2-dioxaborolane) (330 mg, 1.30 mmol, 4.00 eq) were suspended in DMSO (6 mL) and purged with argon for 45 minutes. [1,1′-Bis(diphenylphosphino)ferrocene]dichloropalladium(II)-dichloromethane (53 mg, 0.06 mmol, 0.20 eq) was added and the reaction mixture was heated to 80 °C for 19 hours. After cooling down, water (20 mL) was added and the aqueous phase was extracted three times with dichloromethane (50 mL). The organic phase was dried over MgSO$_4$ and filtered off. The crude product was applied to column chromatography (silica gel, *iso*-hexane:dichloromethane = 2:1 to 1:1) and **5** (182 mg, 0.32 mmol) was obtained as yellow oil in a yield of 99%.

[1]H NMR (300 MHz, CDCl$_3$) $\delta$ = 8.10 (d, *J* = 8.6 Hz, 2H), 7.45 (d, *J* = 8.6 Hz, 2H), 6.97 (d, *J* = 8.7 Hz, 2H), 6.91 (d, *J* = 8.7 Hz, 2H), 2.54 (s, 6H), 1.95 (s, 6H), 1.51 (s, 24H) ppm.

[13]C NMR (76 MHz, CDCl$_3$) $\delta$ = 140.3, 135.8, 135.4, 133.0, 130.8, 129.0, 128.8, 127.6, 126.9, 84.2, 25.3, 22.5, 19.8 ppm. C-B not observed.

HR-MS (MALDI-TOF, dithranol) *m/z*: [M$^+$] calculated for C$_{36}$H$_{44}$B$_2$O$_4$: 561.3460, found: 561.3413, error = -8.37 ppm.

R$_f$ = 0.38 (iso-hexane:dichloromethane = 1:1)

ATR-IR (cm$^{-1}$) *v*: 2976, 2955, 2922, 2854, 1728, 1588, 1568, 1501, 1464, 1447, 1390, 1371, 1331, 1296, 1270, 1213, 1165, 1135, 1109, 1035, 995, 962, 906, 856, 813, 691, 680, 668, 611, 580, 519, 499.

### 3.4. Synthesis of 9,9'-(2,2',6,6'-tetramethyl-[1,1'-binaphthalene]-5,5'-diyl)dianthracene (**2**).[20]

In a 25 mL pressure tube, **5** (56 mg, 0.10 mmol, 1.00 eq), 9-bromoanthracene (307 mg, 1.19 mmol, 12.00 eq) and K$_3$PO$_4$ (507 mg, 2.39 mmol, 24.00 eq) were dissolved in toluene (5 mL), ethanol (1 mL) and water (1 mL) and bubbled with argon for 45 minutes. Pd$_2$(dba)$_3$ (64 mg, 0.07 mmol, 0.70 eq) and DPEPhos (75 mg, 0.14 µmol, 1.40 eq) were added and the reaction mixture was heated to 110 °C for 3 days. After cooling down to room temperature, water (10 mL) was added and the aqueous phase was extracted three times with dichloromethane (30 mL). The combined organic layers were dried over MgSO$_4$, filtered and all volatiles were evaporated under reduced pressure. The residue was purified by column chromatography (silica gel, iso-hexane to iso-hexane:chloroform = 4:1) and recrystallized in ethyl acetate. **2** (26 mg, 0.04 mmol) was obtained as white solid in a yield of 39%.

[1]H NMR (300 MHz, CDCl$_3$) $\delta$ = 8.63 (s, 2H), 8.17 (d, *J* = 3.5 Hz, 2H), 8.14 (d, *J* = 3.6 Hz, 2H), 7.55 − 7.27 (m, 16H), 7.18 (d, *J* = 8.7 Hz, 2H), 6.92 (d, *J* = 8.6 Hz, 2H), 2.08 (s, 6H), 1.93 (s, 6H) ppm.

[13]C NMR (76 MHz, CDCl$_3$) $\delta$ = 135.6, 134.9, 134.6, 133.5, 132.7, 131.8, 131.7, 130.9, 129.1, 129.1, 128.8, 128.76, 126.8, 126.7, 125.9, 125.9, 125.8, 125.4, 20.2, 20.2 ppm.

HR-MS (MALDI-TOF, dithranol) *m/z*: [M$^+$] calculated for C$_{52}$H$_{38}$: 662.2968, found: 662.2990, error = +3.32 ppm.

R$_f$ = 0.43 (iso-hexane:dichloromethane = 3:1)

ATR-IR (cm$^{-1}$) *v*: 3048, 2955, 2921, 2853, 1499, 1440, 1415, 1377, 1336, 1312, 1260, 1217, 1088, 1011, 953, 897, 883, 844, 809, 791, 759, 736, 696, 650, 616, 605, 595, 574, 554, 532, 520, 503, 473, 434.

Mp: 355.0 °C (decomposition).

# 4. NMR spectra

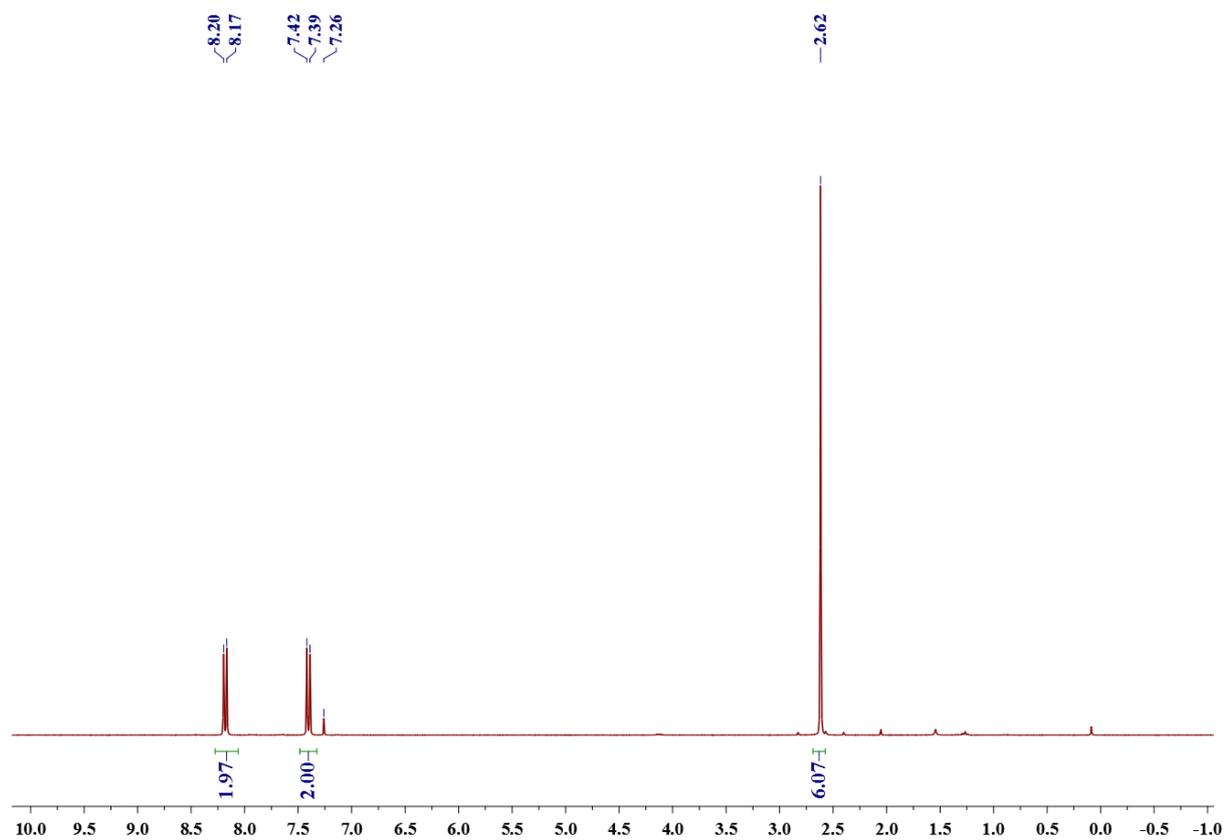

**Figure S9**. ¹H-NMR of **3** dissolved in chloroform-d1, 300 MHz, room temperature.

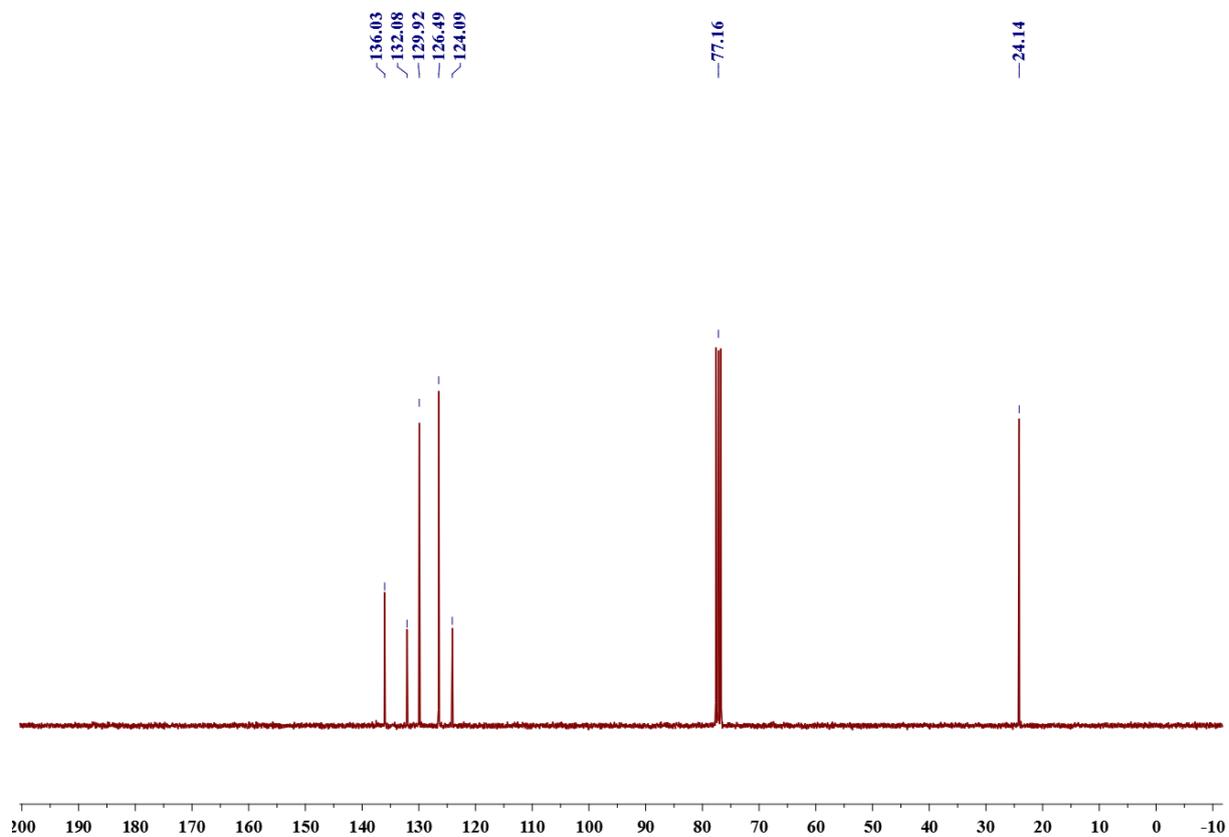

**Figure S10.** ¹³C-NMR of **3** dissolved in chloroform-d1, 300 MHz, room temperature.

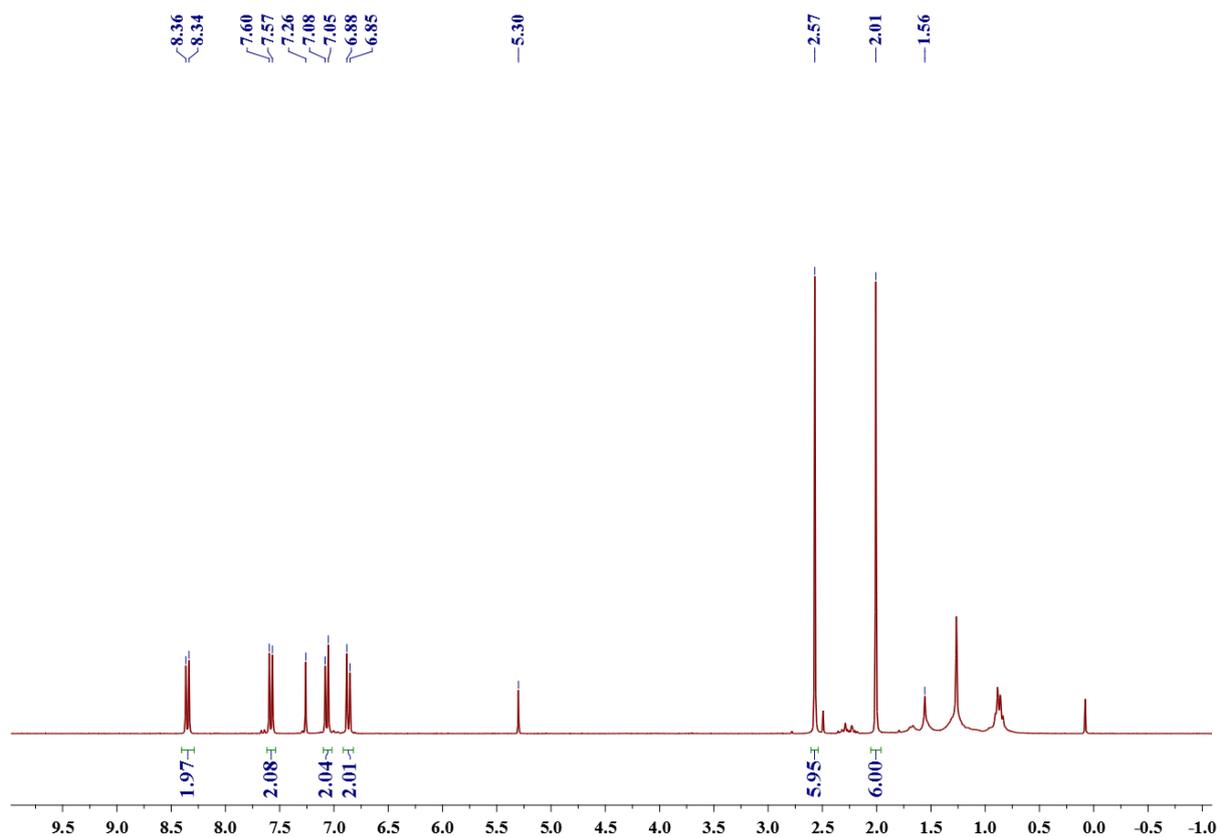

**Figure S11.** ¹H-NMR of **4** dissolved in chloroform-d₁, 300 MHz, room temperature.

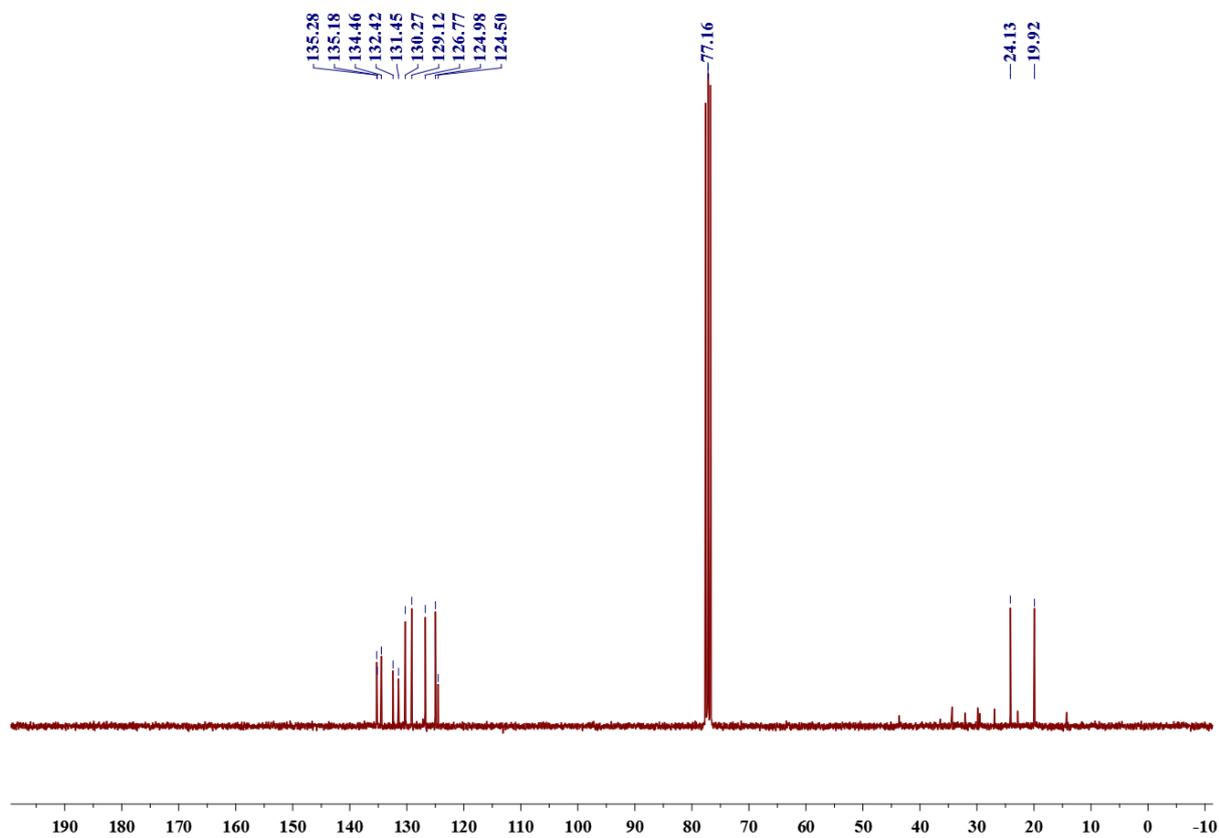

**Figure S12.** ¹³C-NMR of **4** dissolved in chloroform-d₁, 300 MHz, room temperature.

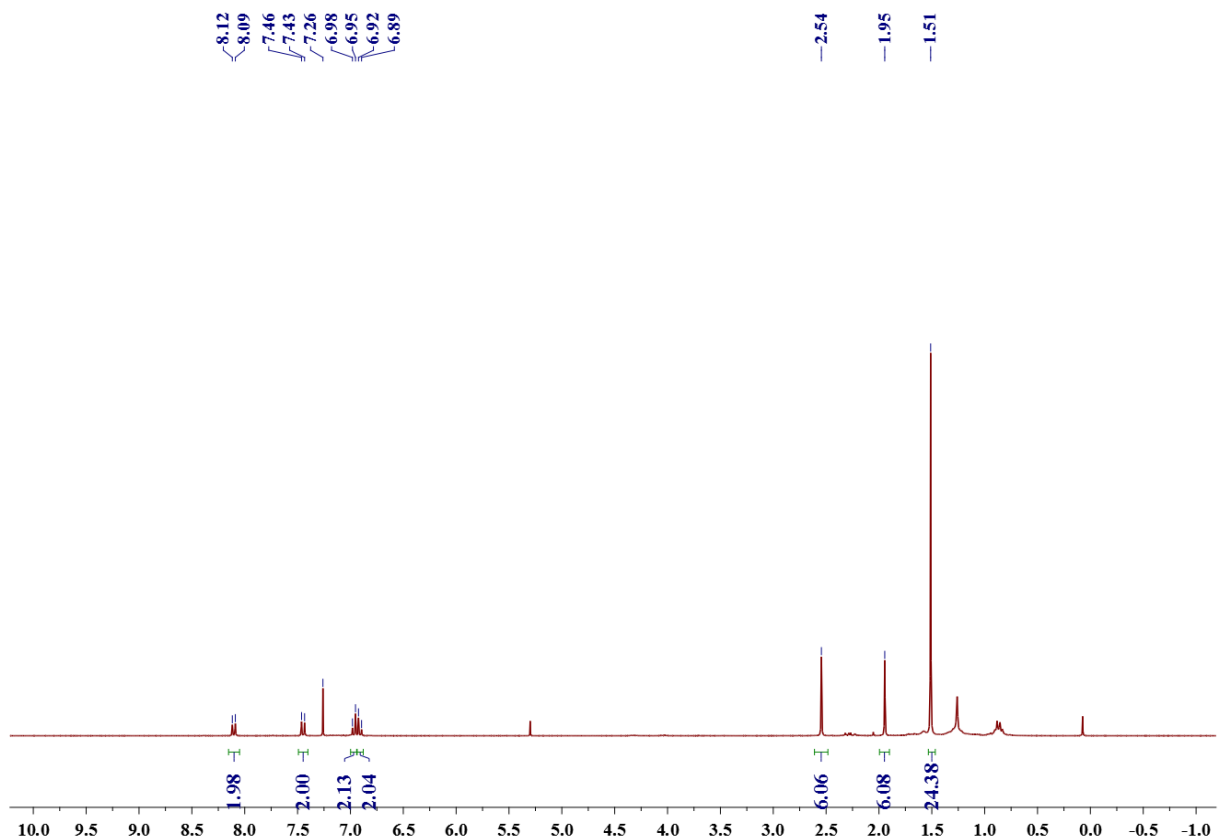

**Figure S13.** ¹H-NMR of **5** dissolved in chloroform-d₁, 300 MHz, room temperature.

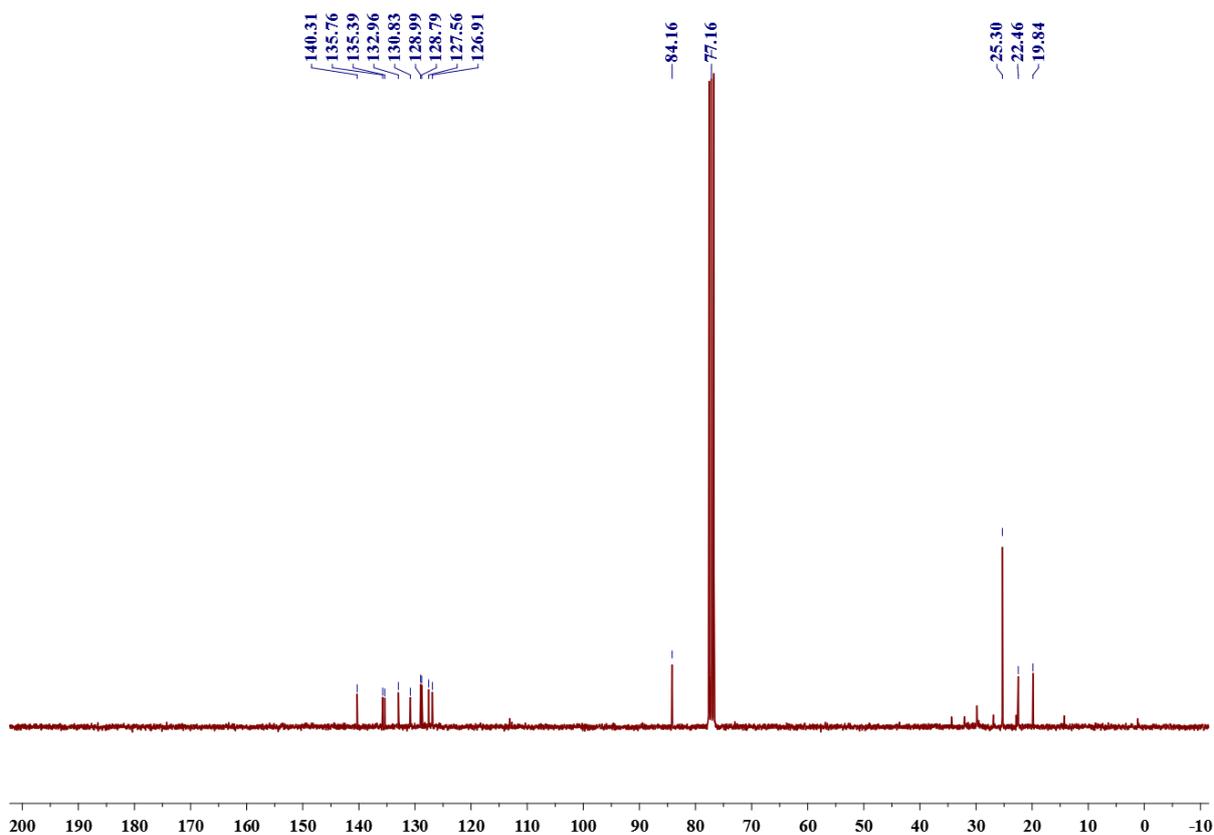

**Figure S14.** ¹³C-NMR of **5** dissolved in chloroform-d₁, 300 MHz, room temperature.

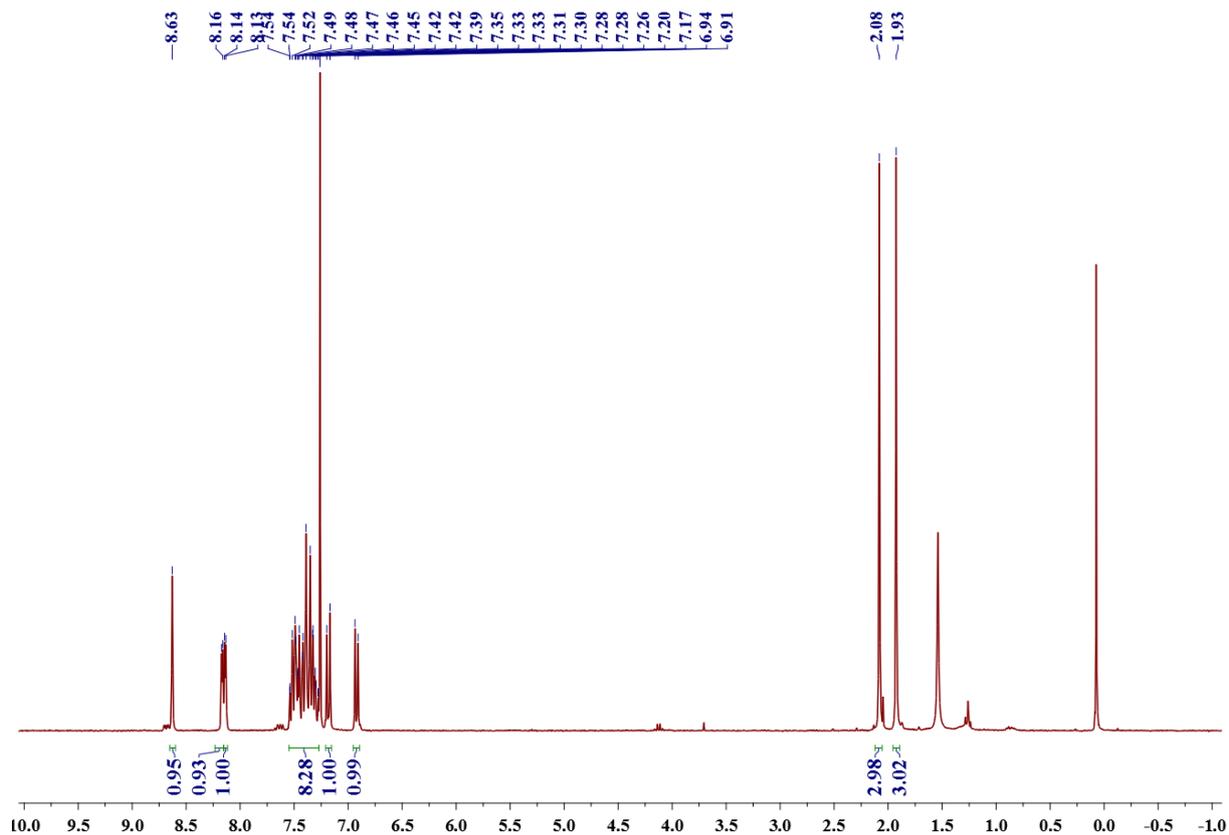

**Figure S15.** ¹H-NMR of **2** dissolved in chloroform-d1, 300 MHz, room temperature.

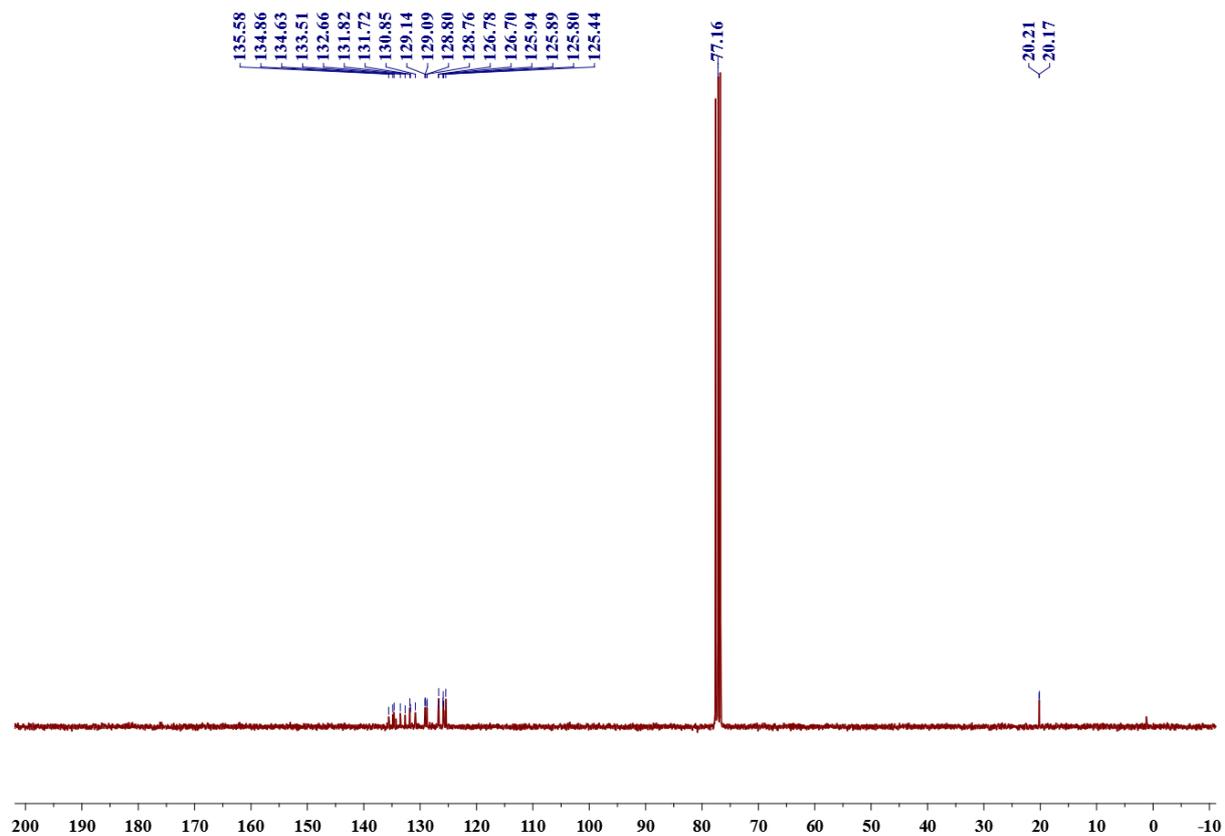

**Figure S16.** ¹³C-NMR of **2** dissolved in chloroform-d1, 300 MHz, room temperature.

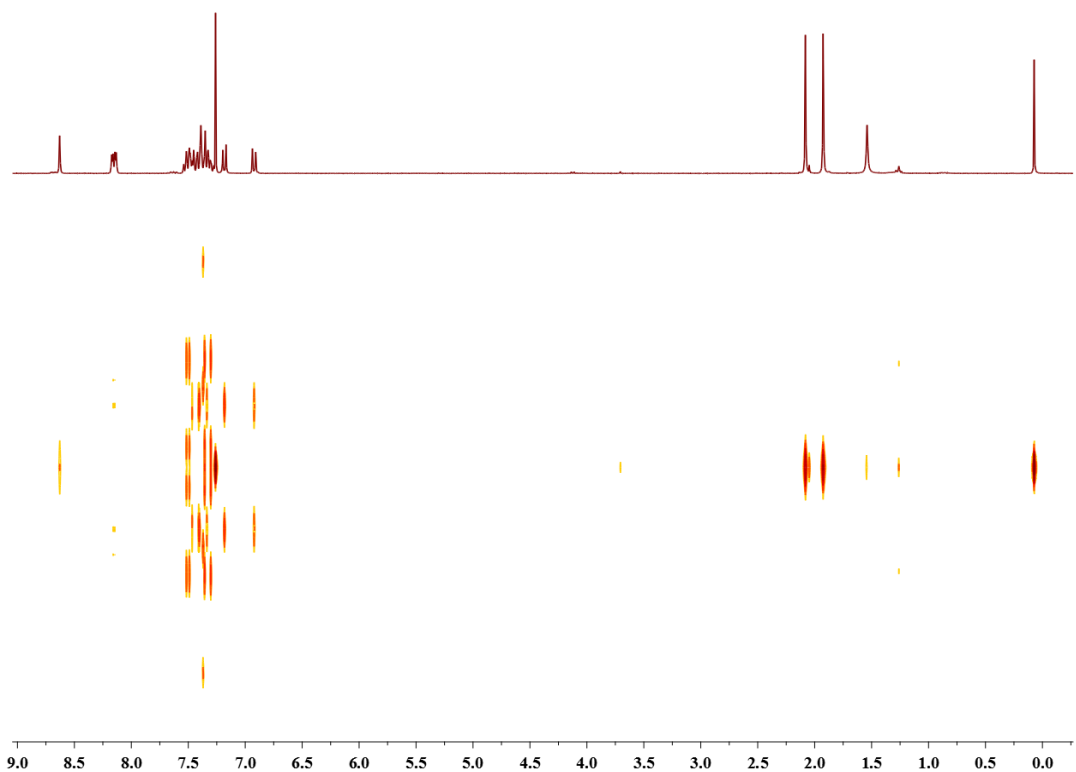

**Figure S17.** JRES-¹H-NMR of **2** dissolved in chloroform-d₁, 300 MHz, room temperature.

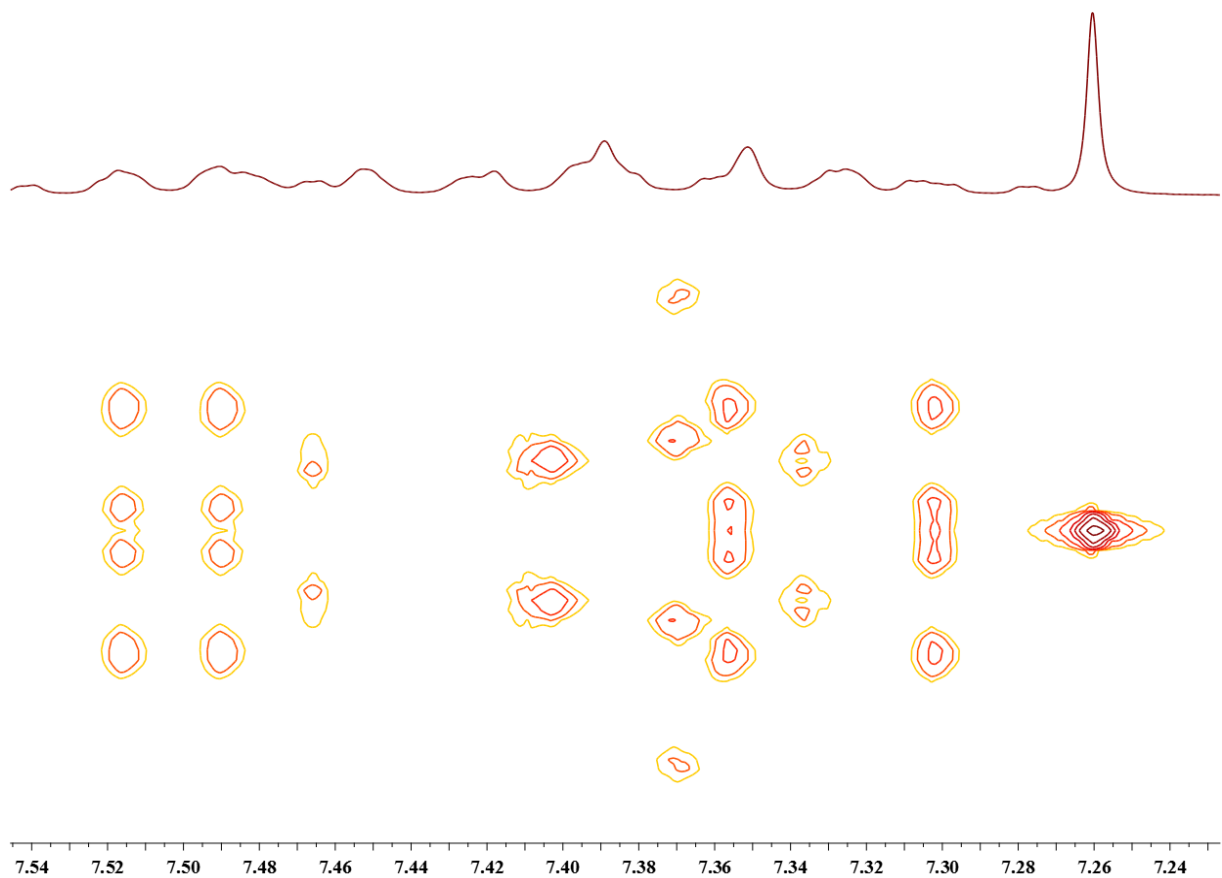

**Figure S18.** JRES-¹H-NMR close-up of **2** dissolved in chloroform-d₁, 300 MHz, room temperature.